\documentclass{kluwer}    
\usepackage{graphicx}
\begin{document}                                                                                   
\begin{article}
\begin{opening}         
\title{Deep Studies of the Resolved Stellar Populations in the Outskirts of M31}
\author{Annette M.~N. \surname{Ferguson}}  
\runningauthor{Annette M.~N. Ferguson}
\runningtitle{Outskirts of M31}
\institute{Kapteyn Institute, PO Box 800, 9700 AV Groningen, The Netherlands}
\date{\today}

\begin{abstract}
We discuss the first results from ongoing studies of the resolved
stellar populations in the outskirts of our nearest large neighbour,
M31.  Deep HST/WFPC2 archival observations are used to construct
colour-magnitude-diagrams which reach well below the horizontal branch
at selected locations in the outer disk and halo, while a panoramic
ground-based imaging survey maps spatial density variations through
resolved star counts to a projected radius of  $\sim 50$kpc.  \end{abstract}
\end{opening} 
     
\section{Introduction} 
The fossil record of star formation and galaxy evolution  is imprinted
on the spatial distribution, ages and metallicities of galactic stellar
populations. Surprisingly little is known about the old and
intermediate-age stellar populations in massive galaxies outside our
own Milky Way.  Located at $\sim$780 kpc, M31 provides our best
opportunity to explore the stellar populations across the face of a
large disk galaxy.  Furthermore, its favourable inclination means that
the disk and halo components at large radii can be easily distinguished
and independently studied.

Most previous studies of M31 have either been based on observations of
single fields (eg. Holland et al 1996, Durrell et al 2001) or on
large-area surveys limited to the bright disk (eg.
Walterbos \& Kennicutt 1988).  It is generally thought that M31
 resembles the Milky Way in many ways, although it is somewhat larger,
more luminous and has a denser stellar halo (see Freeman 1999 for a
review).  The disparity between the mean metallicity of M31's field
halo and that of the Milky Way ($\approx$ factor of 10 lower) was first
recognized by Mould \& Kristian (1986) but remains to this day poorly
understood.

\section{Probing M31 with Deep HST/WFPC2 Pointings}
We searched the HST archive for all deep (T$>$4000s) WFPC2 pointings
towards M31 in the F555W and F814W filters.  We focus here on those
fields which sample the disk and halo at large
radius (R$_{deproj}\gtrsim 20$kpc).

\begin{figure}[H]
\tabcapfont
\centerline{%
\includegraphics[width=8pc]{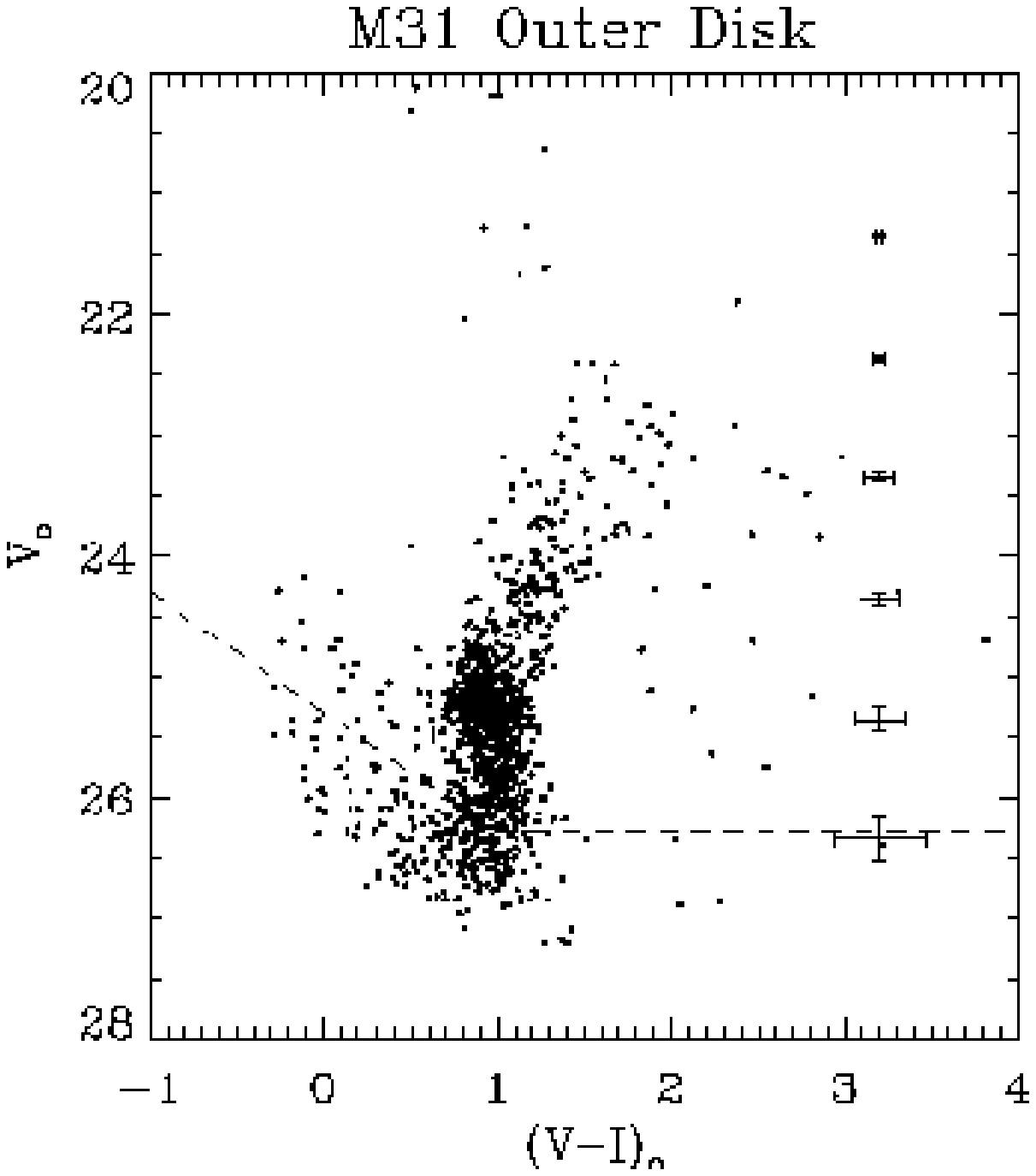} 
\includegraphics[width=8pc]{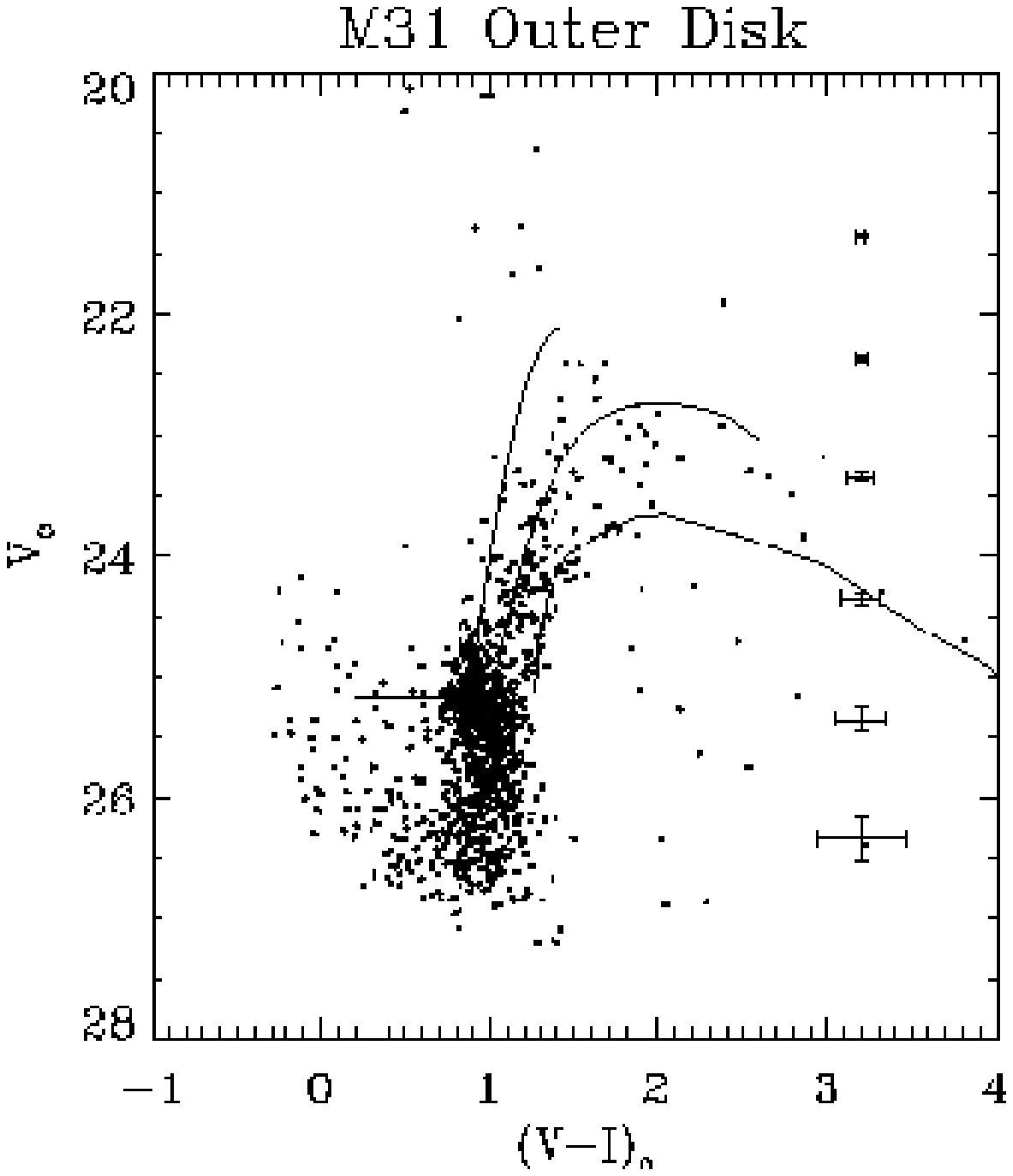}}
\caption{CMDs of the M31 outer disk field (30 kpc or 5 disk
scalelengths). (Left) the dashed line indicates the 75\% completeness
level. (Right) GC fiducials are overlaid corresponding to NGC~6397
([Fe/H]$=-1.9$), 47~Tuc ([Fe/H]$=-0.7$) and NGC~6553 ([Fe/H]$=-0.3$).
Also indicated is the mean V magnitude of the blue horizontal branch
detected in M31 halo by Holland et al (1996).}
\end{figure}

Figure 1 shows the WFPC2 CMD of a field at $\sim30$ kpc (or 5 disk
scale lengths) along the major axis.   Based on extrapolation of the
structural parameters of Walterbos \& Kennicutt (1988), we expect 95\%
of the stars in this field to belong to the disk.  The prominence and
morphology of the red giant branch (RGB) and red clump (RC) indicate an
old-to-intermediate age, fairly metal-rich population.  Comparison of
the outer disk RGB with globular cluster fiducials indicates a mean
metallicity comparable to 47~Tuc ([Fe/H]$=-0.7$).  The significant
width of the RGB exceeds that of photometric errors, and is most easily
explained by an intrinsic dispersion ($\simeq 2$ dex) in the
metallicity of the stellar population. While these properties have been
noted before for the M31 halo (eg. Holland et al 1996), we find here
that they also characterise the outer disk.  See Ferguson 
\& Johnson (2001) for details.

The RC properties, when coupled with the mean metallicity from the RGB
colour, suggest a mean age for the population of $\gtrsim 8$~Gyr
(Cole 1999, Girardi \& Salaris 2001).  This is also consistent with  the
apparent lack of asymptotic giant branch stars above the tip of the RGB
(I$\sim$20.5), which would represent young-to-intermediate age (2--6
Gyr) shell He-burning stars.  Finally,  we note the marginal detection
of horizontal feature in the CMD connecting the blue plume at V$\sim25$ to the RC.  While this could represent the subgiant branch of a
$\sim1$ Gyr population, it is more likely to be an old ($\ge10$ Gyr),
metal-poor ([Fe/H]$\sim-1.7$) horizontal branch (see Figure 1).

Figure 2 shows CMDs for a representative halo-dominated field and a
field in which disk and halo are expected to contribute in roughly
equal amounts.  Apart from stellar density (which reflects both
intrinsic variations as well how much of the WF area was useable in our
analysis), the dominant stellar population in these fields appears
strikingly similar (and indeed, both resemble the outer disk CMD in
Figure 1).    The homogeneity of halo and outer disk populations was
first remarked upon by Morris et al (1994) from shallow ground-based
data, but we show here the similarity holds to well below the
horizontal branch.

\begin{figure}[H]
\tabcapfont
\centerline{%
\includegraphics[width=9pc]{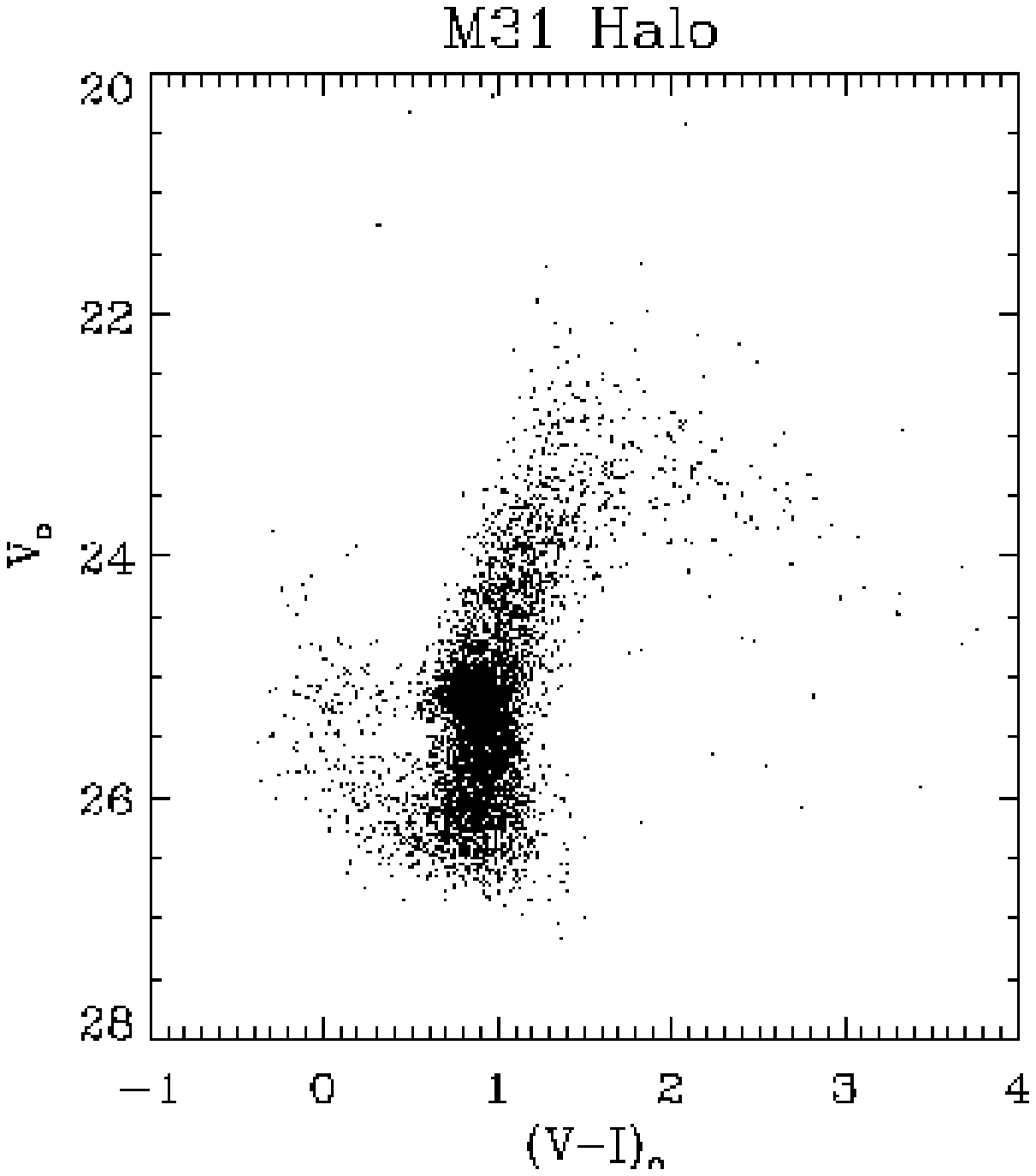} 
\includegraphics[width=9pc]{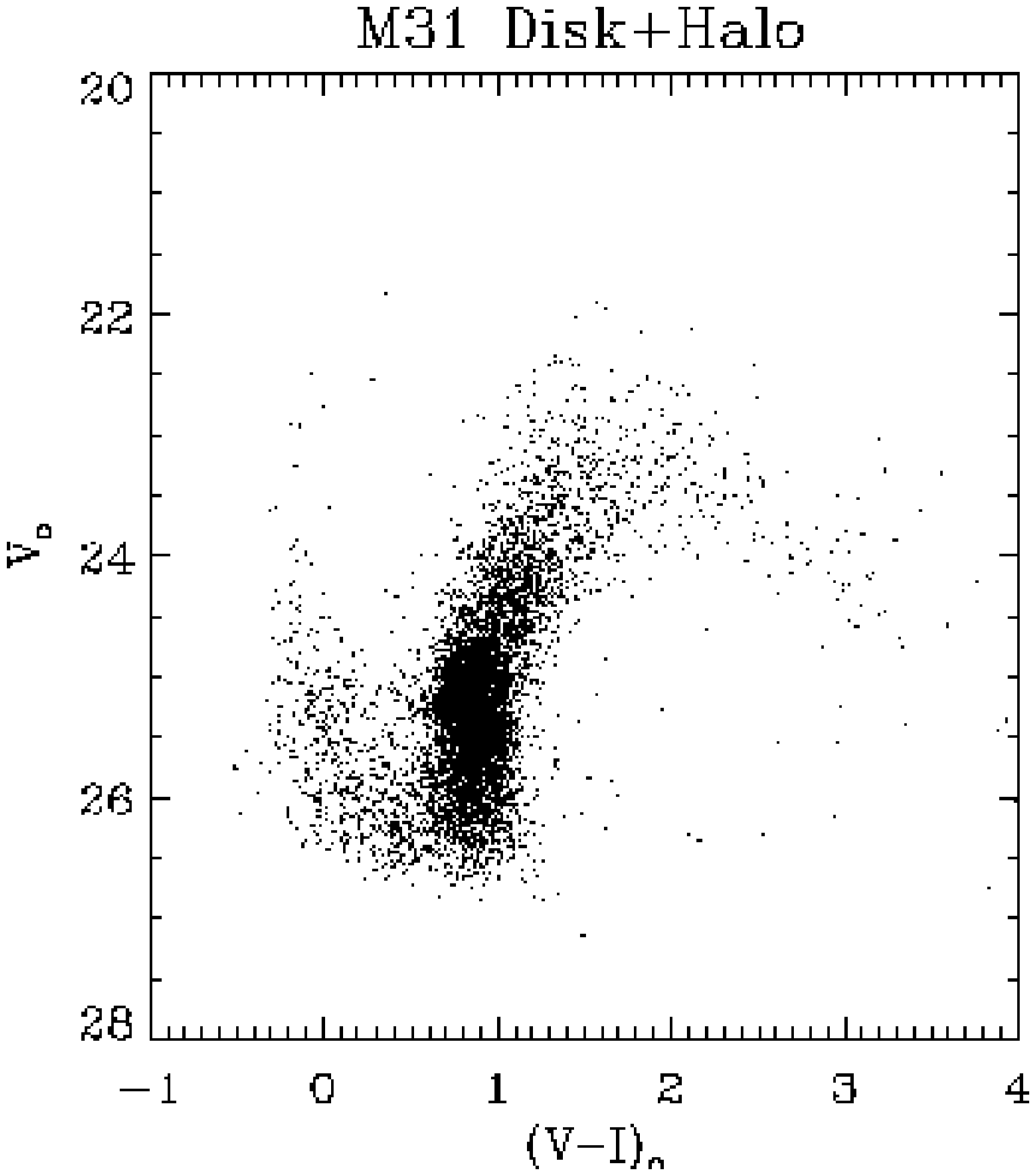}}
\caption{CMDs for a halo-dominated field (left) and a field
which is expected to contain equal numbers of disk and halo
stars (right). Both fields sample roughly the same halo radius (10.5~kpc)
but the disk contribution is greater in the latter.}
\end{figure}

\section{Probing M31 with Panoramic Ground-based Imagery}

While providing detailed information on the stellar populations, the HST
pointings tell us nothing about the large-scale structure of M31
(WFPC2 FOV $\sim 0.3$ kpc$^2$).  We have therefore
embarked upon a ground-based, moderate-depth panoramic imaging survey
of the outskirts of M31 with the INT Wide Field Camera.  To date, 58
contiguous fields (0.3 $\Box^{\circ}$ per pointing) have been observed
in the SE half of the galaxy, mapping out to a projected radius of
$\sim 50$ kpc.

An exciting first result has been the discovery of a giant stream of
stars (overdensity a factor of 2) in the halo of M31 near the southern
minor axis (Ibata et al 2001).  The stream extends beyond the limit of
our imagery, or $\sim40$ kpc at this position angle, and has an average
V-band surface brightness  of $\sim 30\ \rm{mag}/\Box''$.  If an old
coeval population is assumed, the excess stream population is found to
be of similar or slightly higher metallicity than the M31 field halo
and outer disk.

The stream appears to lie along a line which connects M32 and NGC~205,
the two dwarf elliptical companions of M31.  Both these satellites
display rather odd properties for their morphological class --
including young and/or intermediate-age stars and in the case of
NGC~205, cold gas -- and both exhibit distorted isophotes in the outer
regions, suggestive of tidal distortion and possibly disruption.  The
similarity between the metallicities of these dwarfs and that of the
stream suggests that one or both of them may be the origin of the
feature.

\section{Discussion}
We are exploring the stellar populations in the outskirts of M31 using
deep HST archival pointings in combination with a ground-based
panoramic imaging survey.   We summarise our first results as follows: \\
$\clubsuit$~~the disk of M31 appears to have a significant
mean age ($\gtrsim8$~Gyr).  For current cosmologies,  this lookback
time corresponds to a redshift of $\simeq1$ by which about half the
stellar disk was in place at 30~kpc. This may be problematic for theories which
invoke delayed disk formation as a way to circumvent the angular
momentum problem seen in numerical simulations of galaxy formation (eg.
Weil et al 1998).  Equally puzzling is our finding that the dominant
populations in the outer disk and halo are remarkably similar,
suggesting these components have experienced similar formation epochs
and evolutionary histories.\\
$\clubsuit$~~the discovery of a giant stellar stream in the outer halo
of M31 attests to the fact that the hierarchical process of galaxy
formation continues to the present-day and that at least some fraction
of the M31 field halo has not formed {\it in-situ}, but has been
accreted from lower mass objects.  Could this interaction have polluted
the outer regions of M31 to such an extent that it explains that
apparent uniformity of the stellar populations in these parts? Future
observations will help us assess this possibility more throughly.

\acknowledgements
I gratefully acknowledge the input of my collaborators on the
work discussed here---specifically, Rachel Johnson and Nial Tanvir
on the HST studies and Mike Irwin, Rodrigo Ibata, Geraint Lewis and
Nial Tanvir on the INT/WFC survey. Thanks also to the Leids Kerkhoven-Bosscha
Fonds for financial assistance to attend this conference.

\end{article}
\end{document}